\documentclass{iopart}

\usepackage{graphicx}
\usepackage[usenames]{color}
\usepackage{dcolumn}
\usepackage{iopams, amsfonts, amsbsy, amssymb, amsopn, amstext}
\usepackage{bm}
\usepackage{ifpdf}
\ifpdf
\usepackage[colorlinks, plainpages=false, hyperfigures=true]{hyperref}
\usepackage{microtype}
\else
\usepackage[dvipdfm, colorlinks]{hyperref}
\fi

\definecolor{CiteColor}{rgb}{0.250, 0.250, 0.800}
\definecolor{UrlColor} {rgb}{0.741, 0.173, 0.000}
\definecolor{LinkColor}{rgb}{0.000, 0.502, 0.118}
\hypersetup{linkcolor=LinkColor}
\hypersetup{citecolor=CiteColor}
\hypersetup{urlcolor=UrlColor}

\usepackage{xspace}
\newcommand{\pN}{\text{pN}\xspace}

\DeclareMathOperator{\unwind}{unwind}
\newcommand{\Aligned}{\ensuremath{S_{++}^{0.97}}}
\newcommand{\AntiAligned}{\ensuremath{S_{--}^{0.95}}}
\newcommand{\rstar}{\ensuremath{r_{\ast}}}

\newcommand{\MSun}{\ensuremath{M_{\odot}}\xspace}

\renewcommand{\etal}{\textit{et al.}\xspace}

\usepackage{ulem}
\definecolor{darkgreen}{rgb}{0.2,0.7,0.2}

\newcommand{\geoffrey}[1]{#1}
\newcommand{\Mark}[1]{#1}

\newcommand{\geoffreyComment}[1]{}
\newcommand{\mike}[1]{}
\newcommand{\MarkComment}[1]{}

\newcommand{\old}[1]{}
\newcommand{\new}[1]{#1}
\begin{document}

%% \rapid[Accurate gravitational waveforms for BBH mergers with nearly
%% extremal spins]{Accurate gravitational waveforms for
%% binary-black-hole mergers with nearly extremal spins}

\paper[Accurate gravitational waveforms for BBH mergers with nearly
extremal spins]{Accurate gravitational waveforms for binary-black-hole
  mergers with nearly extremal spins}

\author{Geoffrey Lovelace$^1$, Michael Boyle$^1$, Mark A. Scheel$^2$,
  B{\'{e}}la Szil{\'{a}}gyi$^2$} %
\address{$^1$ Center for Radiophysics and Space Research, Cornell
  University, Ithaca, New York, 14853, USA} %
\address{$^2$ Theoretical Astrophysics 350-17, California Institute of
  Technology, Pasadena, California 91125, USA} %

\eads{%
  \mailto{geoffrey@astro.cornell.edu}, %
  \mailto{boyle@astro.cornell.edu}, %
  \mailto{scheel@tapir.caltech.edu}, %
  \mailto{szilagyi@tapir.caltech.edu} %
}

\begin{abstract}
  \new{Motivated by the possibility of observing gravitational waves
    from merging black holes whose spins are nearly extremal (i.e., 1
    in dimensionless units)}, we present numerical waveforms from
  simulations of merging black holes with the highest spins simulated
  to date: (1) a 25.5-orbit inspiral, merger, and ringdown of two
  holes with equal masses and spins of magnitude $0.97$ aligned with
  the orbital angular momentum; and (2) a previously reported
  12.5-orbit inspiral, merger, and ringdown of two holes with equal
  masses and spins of magnitude $0.95$ anti-aligned with the orbital
  angular momentum.  First, we consider the horizon mass and spin
  evolution of the new aligned-spin simulation.
  \new{ During the inspiral, the horizon area and spin evolve in
    remarkably close agreement with Alvi's analytic predictions, and
    the remnant hole's final spin agrees reasonably well with several
    analytic predictions. We also find that the total energy emitted
    by a real astrophysical system with these parameters---almost all
    of which is radiated during the time included in this
    simulation---would be 10.952\% of the initial mass at infinite
    separation.  %This is nearly 3.5 times
    % greater than the emission in the anti-aligned system.
  }%
  Second, we consider the gravitational waveforms for both
  simulations.  After estimating their uncertainties, we compare the
  waveforms to several post-Newtonian approximants, finding
  significant disagreement well before merger, although the phase of
  the TaylorT4 approximant happens to agree remarkably well with the
  numerical prediction in the aligned-spin case.  We find that the
  post-Newtonian waveforms have sufficient uncertainty that hybridized
  waveforms will require far longer numerical simulations (in the
  absence of improved post-Newtonian waveforms) for accurate parameter
  estimation of low-mass binary systems.
\end{abstract}

\date{\today}


% \maketitle

\section{Introduction}
\label{sec:Introduction}
In the next decade, advanced ground-based detectors such as the
advanced Laser Interferometer Gravitational-Wave Observatory (advanced
LIGO)~\cite{Barish:1999,Sigg:2008}, Virgo~\cite{Acernese:2008}, and
the Large-scale Cryogenic Gravitational-wave Telescope
(LCGT)~\cite{Kuroda:2010} are expected to directly observe
gravitational waves for the first time; coalescing black holes are
among the most important sources of gravitational waves for these
detectors.  Numerical predictions of binary-black-hole (BBH) waveforms
are crucial tools for detecting these waves: for example, the
Numerical INJection Analysis (NINJA) project~\cite{NinjaWebPage,ninja}
is testing gravitational-wave search pipelines using numerical BBH
waveforms, and the Numerical-Relativity and Analytical-Relativity
(NR-AR) project~\cite{NRARwebsite} is working to calibrate analytic
template banks for gravitational-wave searches using numerical BBH
waveforms.  Numerical BBH waveforms are also important tools for
parameter estimation~\cite{Hannam:2010,Damour:2010,
  MacDonald:2011ne,Boyle:2011dy}.

Beginning with Pretorius's 2005 breakthrough~\cite{Pretorius2005a},
several groups have successfully completed numerical simulations of
the inspiral, merger, and ringdown of BBHs in a variety of
configurations (see references~\cite{Centrella:2010,McWilliams:2010iq}
for recent reviews). Simulations of BBHs with merging holes whose
spins are nearly extremal (i.e., $\sim 1$ in dimensionless units, the
theoretical upper limit for stationary holes) are a challenging but
potentially important case, since black holes with nearly extremal
spin might exist~\cite{Thorne:1974,GammieEtAl:2004,
  Shapiro:2005,McClintockEtAl:2006,WangEtAl:2006,Rezzolla:2007xa,
  Kesden2008,KesdenEtAl:2010} and thus might be among the BBHs
emitting gravitational waves. Almost all published BBH simulations to
date start with initial data in which 3 of the 4 Einstein constraint
equations are solved analytically using the solutions of Bowen and
York~\cite{bowen79,Bowen-York:1980}; this choice of initial data
limits the black-hole dimensionless spins to $\chi \lesssim
0.93$~\cite{cook90,DainEtAl:2002,HannamEtAl:2009}.  Dain, Lousto, and
Zlochower have closely approached this ``Bowen-York limit'' by
evolving an equal-mass BBH with equal spins of magnitude $\chi=0.924$
aligned with the orbital angular momentum~\cite{DainEtAl:2008}. Note
that the Bowen-York limit is actually far from extremal in terms of
the physical effects of the spin: for example, a black hole with spin
$\chi=0.93$ has only $59\%$ of the rotational energy of an extremal
hole of the same mass.

By using an alternative method to construct BBH initial data, one can
surpass the Bowen-York limit. In reference~\cite{Lovelace2010}, three
of us (Lovelace, Scheel, and Szil{\'{a}}gyi) constructed and evolved
(through 12.5 orbits, merger, and ringdown) BBH initial data (based on
a weighted superposition of two boosted, spinning Kerr-Schild black
holes~\cite{Lovelace2008}) with equal masses and equal spins of
magnitude $\chi=0.94905$ anti-aligned with the orbital angular
momentum.  In this paper, we present a new BBH simulation (through
25.5 orbits of inspiral, merger, and ringdown) with spins of magnitude
$\chi=0.96950$ aligned with the orbital angular momentum. These
simulations are the first to surpass the Bowen-York limit and contain
the most nearly extremal black holes yet simulated, with the black
holes with spin $\chi=0.94905$ and $\chi=0.96950$ having $65\%$ and
$72\%$ as much rotational energy, respectively, as an extremal hole of
the same mass.

In this paper, we consider the gravitational waveforms from these two
simulations.  We begin in Sec.~\ref{sec:methods} by summarizing the
numerical methods we use to construct and evolve rapidly-spinning BBH
initial data and also the methods used to extract and extrapolate the
gravitational waveforms. In Sec.~\ref{sec:AH}, we examine the horizon
mass and spin evolution in the new $\chi=0.96950$ simulation. Then, in
Sec.~\ref{sec:GW}, we examine the emitted gravitational waveforms and
their accuracy. In Sec.~\ref{sec:PNComparison}, we compare the
numerical waves to several post-Newtonian (\pN) approximants. We
conclude with a brief discussion of the implications of our results in
Sec.~\ref{sec:Conclusion}.

\section{Numerical methods}
\label{sec:methods}
\subsection{Initial data}
\label{sec:InitialDataMethods}
\begin{table}
  \lineup
  \begin{tabular}{@{}*{7}{l}@{}}
    \br
    Run & $|1-q|$ & $\chi^z_i$ %%& $M$ 
    & $M_{\text{ADM}}/M$ & $J^z_{\text{ADM}}/M^2$ &
    $M_{\text{final}}/M$ & $\chi^z_{\text{final}}$ \\
    \mr
    $\Aligned$ & $\m1.\times 10^{-8}$ & $+0.96950$ %%& \m1.00024 
    & \m0.992569  & \m1.55877 & \m0.89048  & \m0.94496 \\
    & $\pm6.\times10^{-8}$  & $\pm0.00001$ %%& $\pm 0.00001$ 
    & $\pm0.000009$ &  $\pm0.00003$ & $\pm 0.00002$ & $\pm 0.00001$ \\
    \hline
    $\AntiAligned$ & $\m9.3\times 10^{-6}$ & $-0.949047$ %%& \m0.99987 
    & \m0.99296 & \m0.68404 & \m0.968264 & \m0.3757 \\
    &  $\pm 4.\times10^{-7}$ & $\pm 0.000009$ %%& $\pm 0.00003$ 
    & $\pm 0.00004$ & $\pm 0.00006$ & $\pm 0.000007$ & $\pm 0.0002$ \\
    \br
  \end{tabular}
  \caption{ \label{tab:ID} %
    Some properties of the BBH simulations considered in this
    paper.  The quantity $q$ is the mass ratio, $M$ is the sum of
    the holes' Christodoulou masses, and hole $i$ (where $i=A,B$)
    has dimensionless spin $\chi^z_{i}$ along the $z$ axis (i.e., in
    the direction of the orbital angular momentum).  Also listed
    are the Arnowitt-Deser-Misner (ADM) mass $M_{\text{ADM}}$ and
    angular momentum $J^z_{\text{ADM}}$ (e.g., Eqs.~(25)--(26) of
    reference~\cite{Lovelace2008}) as well as the mass and spin of the
    final black hole.  \new{The quantities $M$, $M_i$, and $\chi_i^z$
      are measured after the initial relaxation and junk radiation
      emission (at times $t=1000 M$ and $t=500 M$ for $\Aligned$
      and $\AntiAligned$, respectively). Note that the mass $M$ and
      spin $\chi^z_i$ are time-dependent; during the inspiral from
      infinite separation to the start of simulations $\Aligned$ and
      $\AntiAligned$, Alvi's formulas~\cite{Alvi:2001mx} predict that
      the total mass $M$ and spin $\chi^z_i$ would change by less than
      one part in $10^6$ and by less than one part in $10^4$,
      respectively.  The ADM quantities $M_{\text{ADM}}$ and
      $J_{\text{ADM}}$ are evaluated at $t=0$, and $M_{\text{final}}$
      and $\chi^z_{\text{final}}$ are measured at the final time of
      the simulation.} All uncertainties are estimated as the
    difference between the quantity at highest and second-highest
    resolution.  %
  }
\end{table}
\begin{table}
  \lineup
  \begin{tabular}{@{}*{5}{l}@{}}
    \br
    Run & $d_0/M$ & $\dot{r}_0/r_{0}$ & $\Omega_0 M$ & $e$ \\ \mr
    $\Aligned$ & 15.362 & \m$0.00084325$ & 0.013815 
    & $6.\times10^{-4} \pm 1. \times 10^{-4}$ \\
    $\AntiAligned$ & 15.368 & $-0.00071390$ & 0.014507 &
    $1.\times10^{-3} \pm 1.\times 10^{-3}$ \\ \br
  \end{tabular}
  \caption{ \label{tab:Ecc} %
    The initial angular velocity $\Omega_0$, 
    radial velocity $\dot{r}_0/r_{0}$, and coordinate separation $d_0$
    and the corresponding estimated orbital eccentricity $e$.  The
    parameters $d_0$, $\dot{r}_0/r_{0}$, and $\Omega_0$ are chosen
    when constructing the initial data; for simplicity, only the first
    five significant figures are shown.  %
  }
\end{table}

To construct BBH initial data with rapid spins, we use the method of
reference~\cite{Lovelace2008} and the references therein: we use a
spectral elliptic solver~\cite{Pfeiffer2003} to solve the extended
conformal thin sandwich equations with quasi-equilibrium boundary
conditions~\cite{York1999,Cook2002,Cook2004,
  Caudill-etal:2006,Gourgoulhon2001,Grandclement2002}. We choose free
data based on a superposition of two boosted, spinning Kerr-Schild
black holes, tuning the freely specifiable parameters with a numerical
root-finding algorithm based on Broyden's method~\cite{Broyden:1965,
  Press2007} to obtain the desired masses and spins. We reduced the
orbital eccentricity using the iterative technique of
reference~\cite{PhysRevD.83.104034} which is based on fits of the
orbital frequency.

We summarize the two configurations we consider in Tables~\ref{tab:ID}
and~\ref{tab:Ecc}.

\subsection{Evolution}
\label{sec:EvolutionMethods}
We evolve the initial data summarized in the
Sec.~\ref{sec:InitialDataMethods} using the Spectral Einstein Code
(SpEC) using the methods summarized in Sec.~III of
reference~\cite{Lovelace2010}, which extend the techniques of
reference~\cite{Szilagyi:2009qz} and the references therein to
accommodate BBHs with spins above the Bowen-York limit.  The evolution
(but not the properties of the resulting gravitational waveform) of
configuration $\AntiAligned$ was first reported in
reference~\cite{Lovelace2010}; we present the evolution of
configuration $\Aligned$ for the first time here.  Full details our
our methods will be given in a future paper; here, we merely summarize
our method, highlighting some of the additional techniques necessary
to merge $\Aligned$.

As described in reference~\cite{Lovelace2010}, we excise the
singularities inside the black holes from our computational domain,
using a time-dependent, adaptively adjusted coordinate mapping to keep
the excision surfaces inside the horizons. Because we do not apply
boundary conditions on the excision surface, the evolution is only
well posed if the excision surface is a pure-outflow surface (i.e, if
it has no incoming characteristic fields).  During the inspiral, we
enforce this condition by controlling the size of the excision surface
such that it tracks the size of the apparent horizon; shortly before
merger (i.e., during the final $\sim 1$ orbit of evolution
$\AntiAligned$ and during the final $\sim 3$ orbits in the evolutions
of $\Aligned$) we control the characteristic speeds directly by
adjusting the velocity of the excision surface. During the final $\sim
0.25$ orbits before merger of $\AntiAligned$ and during the final
$\sim 3$ orbits in $\Aligned$ we employed the spectral adaptive mesh
refinement summarized in reference~\cite{Lovelace2010}. We also note
that when evolving both $\AntiAligned$ and $\Aligned$, we smoothly
change gauge conditions to the damped-harmonic condition described in
reference~\cite{Szilagyi:2009qz} at the beginning of the evolution
instead of shortly before merger.

\begin{figure}
  \includegraphics[width=0.49\textwidth]%
  {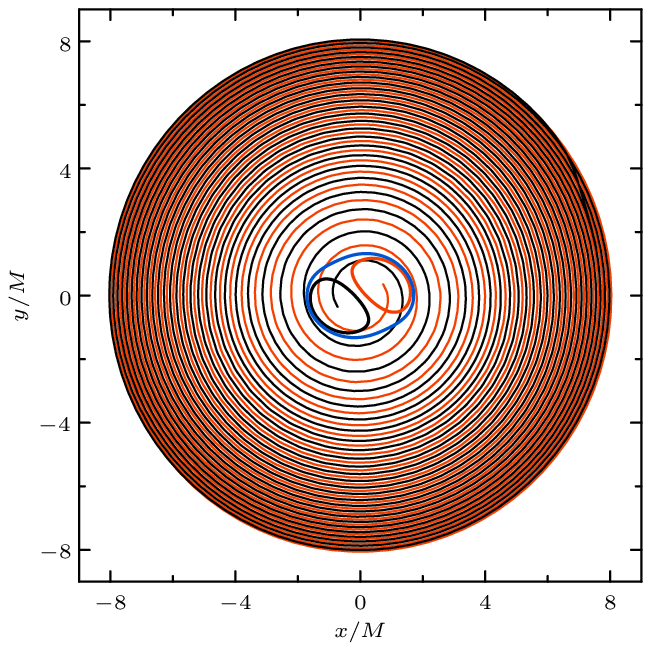}
  \hfil
  \includegraphics[width=0.49\textwidth]%
  {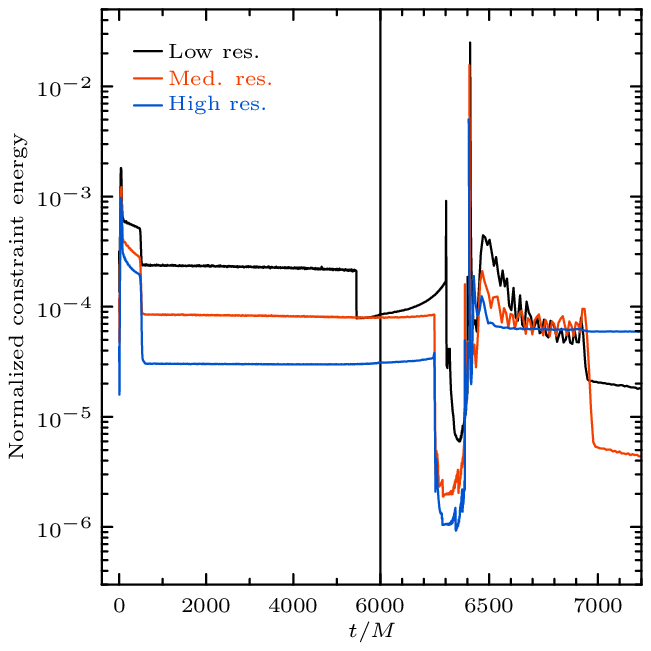}
  \caption{ \label{fig:Align97TrajAndGhCe} %
    The trajectory and constraint violation in simulation $\Aligned$.
    \textit{Left panel:} The trajectories of the centers of the
    individual apparent horizons. Also shown are the individual and
    common horizons at the final time at which the trajectories are
    shown. \textit{Right panel:} For each resolution, the constraint
    violation (specifically, the normalized constraint energy defined
    in Eq.~(71) of reference~\cite{Lindblom2006}) normalized over the
    entire computational domain.  Also note that the horizontal scale
    changes at $t/M=6000$. %
  }
\end{figure}

Because the holes spend more time in a highly dynamical and distorted
state, we find that merging $\Aligned$ requires that our coordinate
mapping must track the apparent-horizon shapes more accurately (i.e.,
to a higher spherical-harmonic resolution $\ell$).  We also find that
we must carefully fine-tune the characteristic speed control to
balance two competing requirements: (1) that the excision surface have
no incoming characteristic fields; and (2) that the excision surface
remain inside the apparent horizon.  Here, we do this fine-tuning
manually in order to merge $\Aligned$; in the future, we plan to
employ a method that handles any tuning automatically. Because the
remnant hole also has a rapid spin [particularly just after it forms
(figure~\ref{fig:Align97SpinVsTime})], we similarly control the
horizon shape and characteristic speeds during the ringdown of
$\Aligned$ (with some manual fine-tuning).

To measure the characteristic speeds on the excision surface with
sufficient accuracy near merger, just before merger in $\Aligned$ we
adopt a computational domain where the individual apparent horizons
lie within a thin, high-resolution spherical shell (instead of within
a set of cylindrical subdomains, as evolution $\AntiAligned$
employed).

The right panel of figure~\ref{fig:Align97TrajAndGhCe} shows the
numerical convergence of our method for the more demanding case
$\Aligned$. The constraints initially grow, then drop as the initial
burst of spurious gravitational radiation leaves the computational
domain. The constraints then remain clearly convergent throughout the
inspiral. Shortly before enabling spectral adaptive mesh refinement,
we found it necessary to increase the resolution of the inner spheres
in the low-resolution run in order to control adequately the
characteristic speeds; this appears as a discontinuous drop in the
low-resolution constraint energy.  As the evolution approaches merger,
the constraint violation grows in spite of the spectral adaptive mesh
refinement.  During ringdown, the constraints rapidly drop as the hole
relaxes to its final state. As the radiation leaves the grid, the
constraints drop sharply in the low and medium resolutions but not in
the high resolution\footnote{\Mark{After a common horizon forms, our
    numerical simulation stops, interpolates onto a new computational
    domain with only a single excised region just inside the common
    horizon, and then continues using this new domain. Thus the
    numerical resolution used during the ``plunge'' (i.e., just before
    merger) is different than during the ``ringdown'' (i.e., just
    after merger).  The lack of convergence at late times in
    figure~\ref{fig:Align97TrajAndGhCe} follows from differences in
    the medium and high plunge resolution. All 3 plunge resolutions
    required different fine-tuning in order to merge; we presume that
    these differences are responsible for the behavior visible in
    figure~\ref{fig:Align97TrajAndGhCe}.  We have verified that for
    the highest plunge resolution, the constraints converge with
    ringdown resolution, even at late times.}}.

\new{Finally, we briefly note the computational cost of these two
  runs.  Because the $\Aligned$ simulation involves higher spins, a
  very large orbital hangup effect (requiring twice as many orbits to
  merge from the same initial separation as the $\AntiAligned$
  simulation), and a large amount of time in a regime where the
  spacetime is highly dynamical, the $\Aligned$ simulation turned out
  to be much more computationally expensive than the $\AntiAligned$
  simulation. Specifically, the high-resolution simulation $\Aligned$
  required $\approx 110,000$ cpu hours ($\approx 120$ days of
  wallclock time). For comparison, the high-resolution $\AntiAligned$
  simulation required $\approx 20,000$ cpu hours ($\approx 20$ days of
  wallclock time).}

\subsection{Waveform extraction}

We extract the gravitational waveform $h$ using the
Regge-Wheeler-Zerilli formalism~\cite{ReggeWheeler1957, Zerilli1970b,
  Sarbach2001, Rinne2008b} on concentric spheres with radii from
roughly $r=100\,M$ to $400\,M$; we then
extrapolate\footnote{\geoffrey{We extract at finite radii and then
    extrapolate, since our computational domain only extends out to
    some finite radius.  Extrapolation is meant to eliminate
    near-field effects and gauge dependence; to verify that the
    extrapolated waves do not retain any residual gauge dependence,
    the extrapolated waves could be compared with waveforms obtained
    using Cauchy Characteristic Extraction (CCE; see
    reference~\cite{Babiuc:2010ze} and the references therein for
    details).  In the future, we plan to compare CCE waveforms with
    the extrapolated waveforms presented in this paper.}}  
the waveforms to
infinite radius using the method of Boyle and
Mrou{\'{e}}~\cite{Boyle-Mroue:2008} with polynomials of order $N=4$.
The waveforms are decomposed in the standard way~\cite{Brown2007} as
modes $h_{\ell,m}$ of spin $s=-2$ spherical harmonics.  We use these
modes to define the amplitude and phase of the waveforms in the usual
way:
\begin{equation}
  \label{eq:AmpPhaseDef}
  A_{\ell,m}(\tau) = \left| r\, h_{\ell,m}(\tau) / M \right| \qquad
  \phi_{\ell,m}(\tau) = \unwind \left\{ \arg[ h_{\ell,m}(\tau)]
  \right\}~,
\end{equation}
where the factor of $r/M$ removes the radial dependence from
$A_{\ell,m}$ and the $\unwind$ function removes discontinuities of
$2\pi$ in the data caused by branch cuts~\cite{Boyle2007}.  We also
use the frequency $\omega_{\ell,m} (\tau) = \partial_{t}\,
\phi_{\ell,m}(\tau)$.  We will occasionally drop the subscripts on
these quantities, implicitly referring to $(\ell,m) = (2,2)$.

\section{Mass and spin evolution in the $\Aligned$ simulation}
\label{sec:AH}

\begin{figure}
  \includegraphics[width=0.49\linewidth]%
  {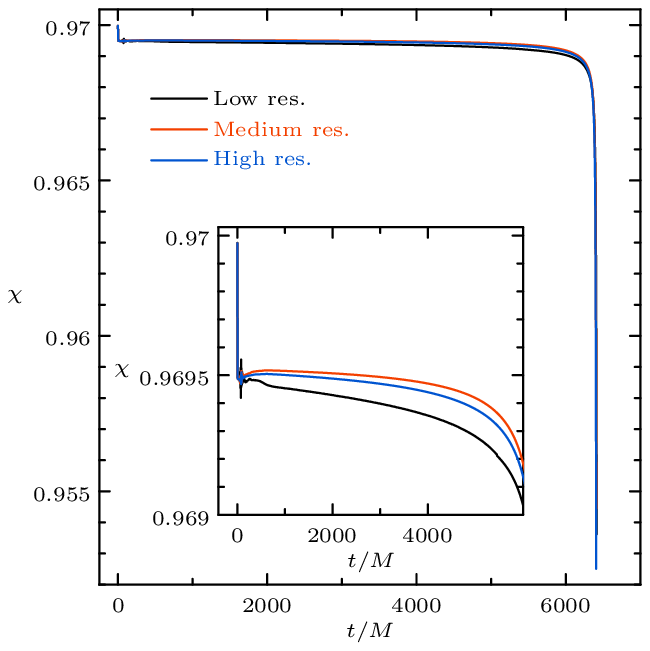} \hfil
  \includegraphics[width=0.49\linewidth]%
  {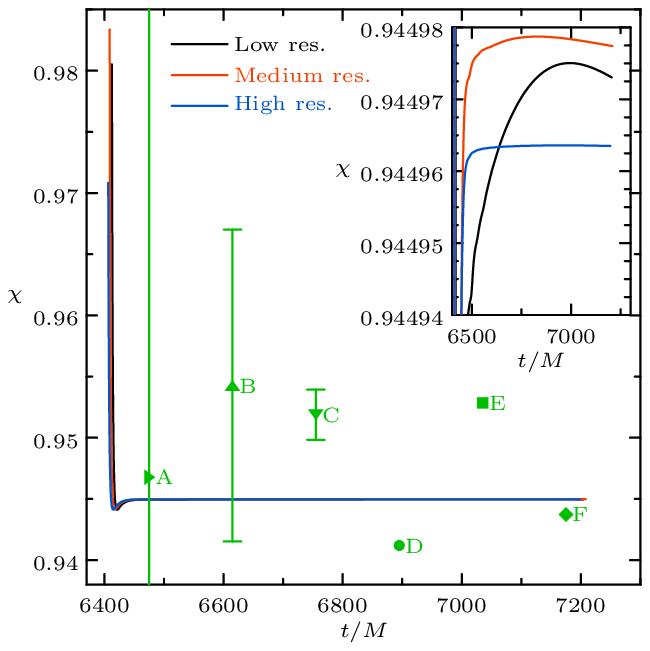}
  \caption{ \label{fig:Align97SpinVsTime} %
    The spin evolution in simulation $\Aligned$.  \textit{Left panel:}
    The dimensionless spin $\chi$ as a function of time for one of the
    individual horizons. The inset zooms in on the spin during the
    inspiral. \textit{Right panel:} The dimensionless spin $\chi$ of
    the common horizon as a function of time. Also shown, as
    individual data points with error bars where applicable, are the
    final spin predictions given by the analytic formulas in
    references \cite{Tichy2008}~(``A''), \cite{Barausse2009}~(``B''),
    \cite{Rezzolla:2007xa}~(``C''), \cite{Campanelli2006c}~(``D''),
    \cite{Buonanno2008a}~(``E''), and \cite{Boyle2007b}~(``F''). The
    inset zooms in on the final spin of the common horizon.  The
    predictions are shown at different times simply for convenient
    separation. %
  }
\end{figure}

In figure~\ref{fig:Align97SpinVsTime}, we show the dimensionless spin
as a function of time for each resolution of the $\Aligned$
simulation.  During the initial relaxation, the holes absorb energy,
causing the dimensionless spin to quickly relax from $\chi=0.9700$ at
time $t=0$ to $\chi=0.9695$ at time $t=1000 M$.

Similarly to reference~\cite{DainEtAl:2008}, we observe a very large
orbital hangup~\cite{Campanelli2006c, Rezzolla:2007rd, Hannam2007c}
during the long inspiral: starting from the same initial coordinate
separation (and thus at approximately the same initial orbital
frequency), case $\Aligned$ requires more than twice as many orbits to
merge as does case $\AntiAligned$ (compare the left panel of
figure~\ref{fig:Align97TrajAndGhCe} and the top-left panel of figure~3
in reference~\cite{Lovelace2010}) and reaches roughly twice the
orbital frequency.  During this long inspiral, the spin remains above
$\chi=0.969$ during the first 21.5 orbits but then decreases near
merger as the spin angular momentum is transformed into orbital
angular momentum via tidal interactions.

\begin{figure}
  \includegraphics[width=0.495\linewidth]{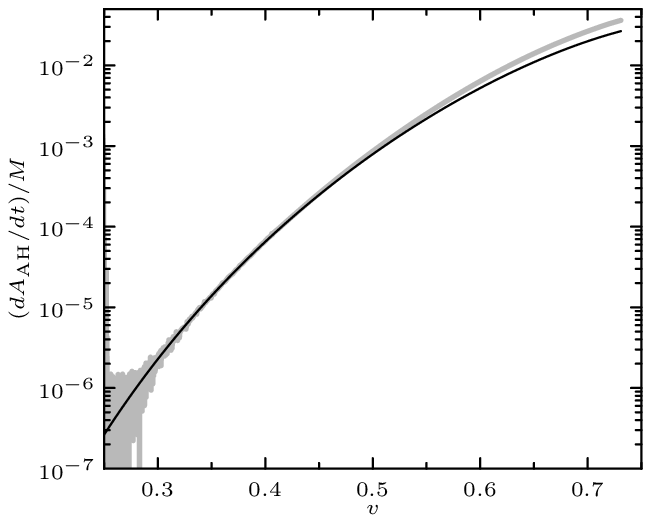} %
  \hfil
  \includegraphics[width=0.495\linewidth]{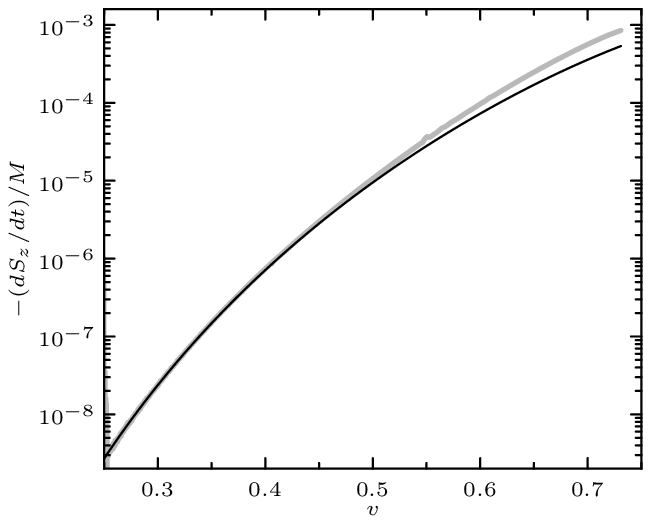}
  \caption{ \label{fig:AlviExpressions} %
    Change in irreducible mass and angular momentum of the horizons in
    the $\Aligned$ system compared with Alvi's
    predictions~\cite{Alvi:2001mx}.  \textit{Left panel:} rate of
    change of irreducible (apparent-horizon) mass of the individual
    horizons.  \textit{Right panel:} rate of change of spin of the
    individual horizons.  In each case, the noisier (thick gray) line
    represents data measured directly in the simulation, while the
    smoother (thin black) line is given by Alvi's equation~(11).  The
    horizontal axis is parameterized by $v = (M\, \Omega)^{1/3}$,
    where $\Omega$ is the orbital frequency measured in the
    simulation.  Note that $v \gtrsim 0.5$ corresponds to the last
    $\sim 40\, M$ of time before merger. %
  }
\end{figure}

The mechanism by which this transformation of angular momentum takes
place has been described by numerous authors~\cite{Teukolsky,
  Hartle:1973zz, Hartle:1974, Price:1986yy, Thorne-Price-MacDonald}
including Alvi~\cite{Alvi:2001mx}, who gave expressions for the rate
at which energy and angular momentum would be transferred in
comparable-mass binaries.  In figure~\ref{fig:AlviExpressions} we plot
those rates as measured in the $\Aligned$ simulation and compare to
those predicted by Alvi.  Alvi's expression uses the post-Newtonian
velocity parameter $v$ which we set to $v = (M\, \Omega)^{1/3}$, where
$\Omega$ is the orbital angular frequency measured in the simulation.
Though this comparison is gauge dependent, we find very good
agreement---within the numerical uncertainty until very late in the
simulation.  A similar comparison for the $\AntiAligned$ case is not
as useful because the numerical uncertainties are far larger, though
the result is consistent within the larger uncertainties.  Note that
this transfer of angular momentum is a 2.5-\pN spin effect which is
incorporated into the calculation of the \pN waveforms in
section~\ref{sec:PNComparison}.

% \begin{figure}
%   \includegraphics[width=0.495\linewidth]{dMdt_UU_Lev6} %
%   \hfil
%   \includegraphics[width=0.495\linewidth]{dSdt_UU_Lev6}
%   \caption{ \label{fig:AlviExpressions} %
%   Energy and angular momentum flow through the horizons in the
%   $\Aligned$ system compared with Alvi's
%   predictions~\cite{Alvi:2001mx} based on the orbital frequency of
%   the binary.  \textit{Left panel:} rate of change of Christodoulou
%   mass of the individual horizons.  \textit{Right panel:} rate of
%   change of spin of the individual horizons.  In each case, the
%   noisier (blue) line represents data measured directly in the
%   simulation, while the smoother (green) line is given by Alvi's
%   equation~(11). %
% }
% \end{figure}

When the common apparent horizon first forms, its spin is very nearly
extremal but then quickly relaxes as the common horizon expands,
eventually settling to a final spin $\chi_{\text{final}}=0.94496 \pm
0.00001$, which is roughly consistent with the predictions of several
analytic approximations (right panel of
figure~\ref{fig:Align97SpinVsTime}).  The mass of the final hole is
$M_{\text{final}}/M=0.89048 \pm 0.00002$.  Alvi's formulas suggest
that the mass would change by less than a part in $10^{6}$ prior to
the beginning of our simulation if the binary had inspiraled from
infinite separation; therefore, a ``real'' binary would have radiated
$E_{\rm rad}/M = 1-M_{\text{final}} / M = 10.952\% \pm 0.002\%$ of its
initial mass \new{($=12.299\% \pm 0.003\%$ of its final mass)
  throughout its entire inspiral, merger, and ringdown. This
  efficiency is comparable to that of a supernova ($\approx 15\%$ of
  the final core mass radiated; see, e.g., equation~(18.1.1) of
  reference~\cite{shapiro83}) but corresponds to a larger total energy
  radiated (since the mass of a BBH is typically larger than the final
  core mass after a supernova).}
For comparison, the simulation in reference~\cite{Scheel2009} implies
that an equal-mass \emph{nonspinning} binary system would radiate
about 5\% of its mass\new{, while table~\ref{tab:ID} shows that the
  $\AntiAligned$ system radiates about 3.17\% of its mass.  When the
  merger occurs at a frequency in the sensitive band of a
  gravitational-wave detector, we can therefore expect that an
  aligned-spin system should have significantly larger SNR than a
  similar system with anti-aligned spins.}  

\section{Gravitational waveforms and post-Newtonian comparisons}

To compare two waveforms, $A$ and $B$, we need to align them by fixing
the arbitrary relative time and phase offsets.  Here $A$ and $B$ may
refer to two numerical waveforms with different resolutions or
extrapolation orders, or $A$ and $B$ may refer to \pN and numerical
waveforms.  Following reference~\cite{Boyle2008a}, we align the
waveforms by minimizing the difference in their phases over a certain
range.  Specifically, we minimize\footnote{For fixed $\Delta t$, the
  optimal $\Delta \phi$ can be obtained analytically, reducing the
  minimization to a one-dimensional problem.  See
  reference~\cite{Boyle2008a} for details.} the quantity
\begin{equation}
  \label{eq:AlignmentMinimization}
  \Xi(\Delta t, \Delta \phi) = \int_{t_{1}}^{t_{2}}\, \left[
    \phi_{A}(t) - \phi_{B}(t-\Delta t) - \Delta \phi \right]^{2}\,
  \rmd t~.
\end{equation}
Each mode of waveform $B$ is then transformed as
\begin{equation}
  \label{eq:PNTransformation}
  h_{\ell,m}(t) \to h_{\ell,m}(t+\Delta t)\, \rme^{-\rmi\, m\, \Delta
    \phi/2}~.
\end{equation}
Note that $\phi$ refers to the phase of the $(\ell,m) = (2,2)$ mode
only; the values of $\Delta t$ and $\Delta \phi$ are determined once,
then each mode is transformed by this equation.

The optimal values of $\Delta t$ and $\Delta \phi$ determined by
minimizing equation~\eref{eq:AlignmentMinimization} clearly depend on
the range of integration $(t_{1}, t_{2})$.  We choose that range based
on the frequencies of the waveform~\cite{MacDonald:2011ne} so that
$\omega(t_{1}) \approx 0.033$ and $\omega(t_{2}) \approx
0.038$.\footnote{This range corresponds to $\delta\omega / \omega
  \approx 15\%$, which is somewhat larger than the 10\% minimum
  recommended by MacDonald \etal~\cite{MacDonald:2011ne}.  We use a
  larger range to ensure that the alignment is not skewed by small
  oscillations in the data.}  This gives us a common basis for
comparison of the ranges used in the two cases of aligned and
anti-aligned spins, despite the very different lengths of time over
which they inspiral.

\subsection{Waveform accuracy}
\label{sec:GW}
\begin{figure}
  \includegraphics[width=0.495\linewidth]%
  {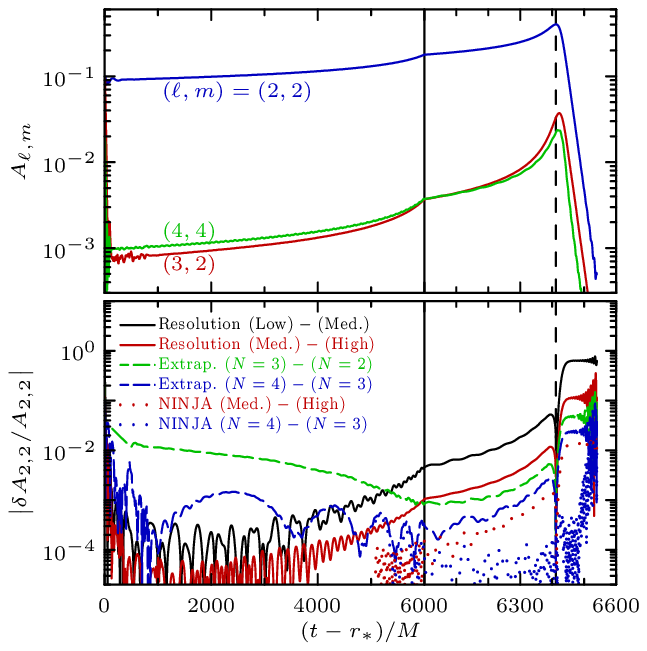}
  \hfil
  \includegraphics[width=0.495\linewidth]%
  {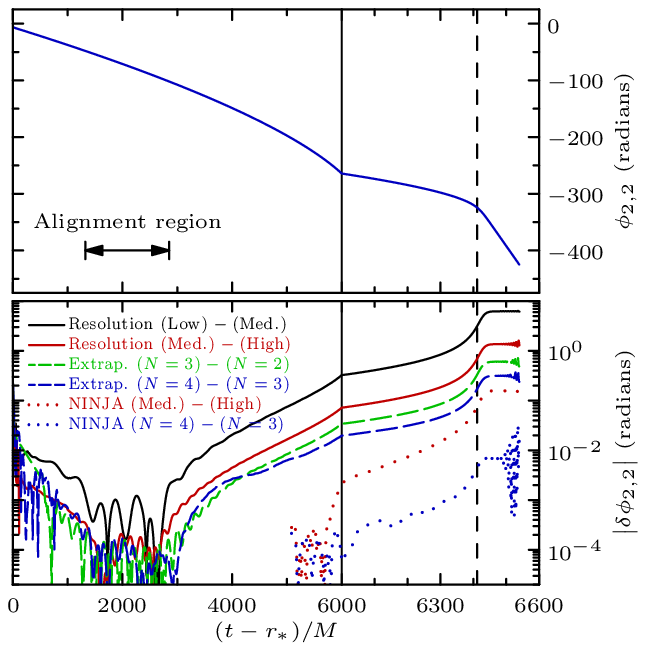}
  \caption{ \label{fig:Align97GW} %
    The extrapolated gravitational-wave amplitudes and phase for
    simulation $\Aligned$.  \textit{Left panel:} The dominant wave
    amplitudes $A_{2,2}$, $A_{3,2}$, and $A_{4,4}$ at high resolution
    (top) and relative differences $|\delta A_{2,2}/A_{2,2}|$ between
    resolutions or between extrapolation orders (bottom).
    \textit{Right panel:} The phase $\phi$ at high resolution (top)
    and differences $|\delta \phi|$ between resolutions (bottom). When
    computing differences, the waveforms are aligned as in
    \eref{eq:AlignmentMinimization} between $(t-\rstar)/M = 1322$ and
    $(t - \rstar)/M = 2852$.  The merger time $(t-\rstar)/M = 6411$ is
    the time at which the $(2,2)$ amplitude is maximal, denoted by the
    vertical dashed lines. If the waveform were instead truncated to 5
    orbits before merger (the NINJA-2 length requirement), the
    amplitude and phase errors would drop significantly (dotted
    lines).  Note that the scale on the horizontal axis changes at
    $(t-\rstar)/M = 6000$ in each plot for improved visibility of the
    merger and ringdown. %
  }
\end{figure}

\begin{figure}
  \includegraphics[width=0.495\linewidth]%
  {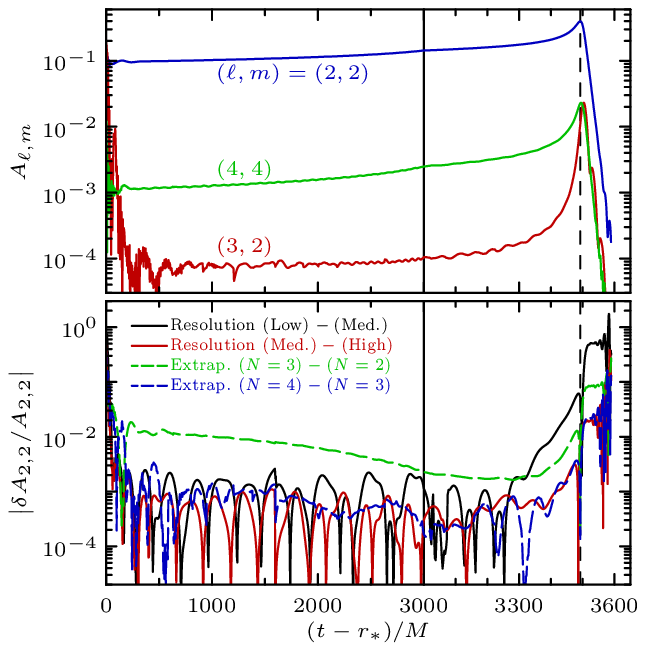}
  \hfil
  \includegraphics[width=0.495\linewidth]%
  {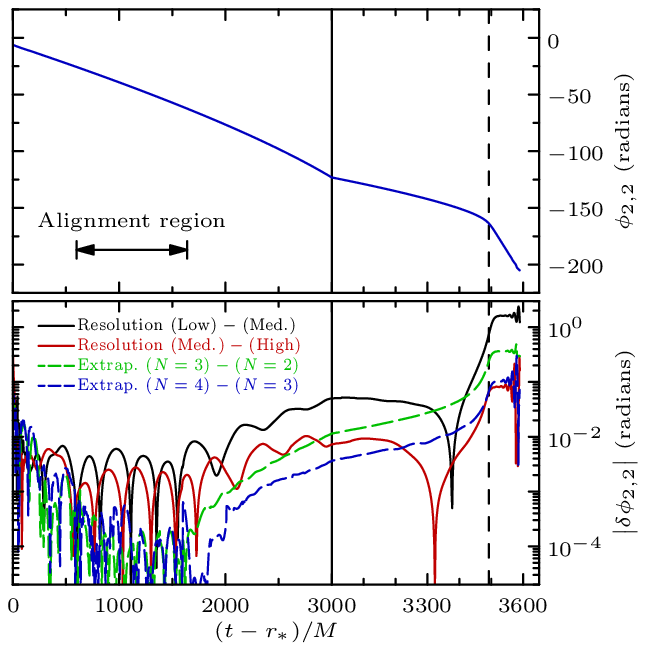}
  \caption{ \label{fig:Anti95GW} %
    The extrapolated gravitational-wave amplitudes and phase 
    as in figure~\ref{fig:Align97GW} but
    for the $\AntiAligned$ simulation.  
    Here, alignment is performed between $(t-\rstar)/M =
    600$ and $(t-\rstar)/M = 1640$.  This simulation is much shorter
    than the aligned case (about half the number of orbits), even
    though both simulations start at the same coordinate separation
    and consequently at roughly the same frequency.  Here the
    anti-aligned spins cause the holes to merge more quickly than an
    analogous non-spinning binary, whereas in case $\Aligned$ the
    binary experiences a large orbital hangup.  Note that the scale on
    the horizontal axis changes at $(t-\rstar)/M = 3000$ in each plot
    for improved visibility of the merger and ringdown. %
  }
\end{figure}

We plot the amplitudes of the three dominant modes and the phase of
the dominant $(\ell, m) = (2,2)$ mode in the upper panels of
figure~\ref{fig:Align97GW} (for $\Aligned)$ and
figure~\ref{fig:Anti95GW} (for $\AntiAligned$).  We also estimate the
accuracy of the waveforms by measuring convergence with respect to
increasing numerical resolution in the simulations and with respect to
increasing order of the polynomial used for extrapolation to infinite
radius.  The relative amplitude convergence and phase convergence are
plotted in the lower panels of the two figures.

The overall uncertainty estimate for a given quantity is the sum of
the absolute values of the resolution convergence and the
extrapolation convergence.  That is, in the notation of the figures,
we estimate
\begin{equation}
  \label{eq:WaveformUncertainty}
  \text{Uncertainty} \approx \left| (\text{Med.}) - (\text{High})
  \right| + \left| (N=4) - (N=3) \right|~.
\end{equation}
For a waveform to be included in the NINJA-2 data
set~\cite{NinjaWebPage}, the amplitude and phase of the $(2,2)$ mode
must be accurate at merger to within $5\%$ and $0.5\, \text{rad}$,
respectively.  The $\AntiAligned$ case exceeds these requirements.
The $\Aligned$ case, however, exceeds the amplitude requirement but
does not meet the phase accuracy requirement.  Through the time of
merger, the amplitude uncertainty never exceeds $2\%$, but the phase
uncertainty at the time of merger is $0.9\, \text{rad}$.

While the complete $\Aligned$ waveform does not meet the NINJA-2
\emph{accuracy} requirement for phase, this simulation's $25.5$ orbits
far exceed the NINJA-2 \emph{length} requirement of 5 orbits aligned
with \pN before a frequency of $M\, \omega = 0.075$.  Indeed, if we
omit the first \new{$5000\,M$ of the $\Aligned$ waveforms, the
  simulation still easily meets the length requirements.  Evaluating
  the errors by aligning between the beginning of the shortened
  waveforms and $M\, \omega = 0.075$, the convergence measure improves
  by a factor of 10, and} the simulation comfortably exceeds the
NINJA-2 accuracy requirements with a phase uncertainty of $0.1\,
\text{rad}$ at the time of merger, as shown by the dotted lines in the
lower panels of figure~\ref{fig:Align97GW}.  Thus, we see that using
accuracy requirements of this sort without regard to the length of the
simulation actually creates a perverse incentive to produce
\emph{shorter} numerical waveforms, which \emph{decreases} the
accuracy of complete ``hybridized'' waveforms~\cite{Boyle:2011dy}.  An
exactly analogous situation occurs when the criteria require a
particular match with respect to a detector noise curve but do not
stipulate a mass at which that match should be measured; again, the
incentive is to produce shorter waveforms.

\subsection{Comparison with post-Newtonian approximations}
\label{sec:PNComparison}
The post-Newtonian waveform is constructed in two steps: (1)
computation of the orbital phase of the binary; and (2) computation of
the amplitude of the waveform using that phase.  For nonspinning
systems, the formulas needed for those computations have been
calculated to 3.0-\pN order beyond the leading order in amplitude and
3.5-\pN beyond leading order in phase.  Additional spin-orbit and
spin-spin terms are available to 3.0-\pN order in phase and
amplitude,\footnote{See reference~\cite{BlanchetEtAl:2011} and
  references therein.} though not all of these have yet been expressed
in a useful form for generating waveforms.  We use the expressions
given in appendix 1 of~\cite{Brown2007} for the flux, orbital energy,
and tidal heating; and the expressions given in equation (9.4) of
reference~\cite{BFIS} and appendix 2 of reference~\cite{Brown2007} for
the waveform amplitudes.  The sole addition we make is the inclusion
of a recently published spin-orbit contribution to the flux.
In~\cite{Brown2007}, equation (A.13) should be supplemented by adding
a term~\cite{BlanchetEtAl:2011} as
\begin{equation}
  \label{eq:PNEnergyAddition}
  \mathcal{F}(v) \to \mathcal{F}(v) + \frac{32}{5}\, v^{10}\,
  \eta^{2}\, \left\{v^{6}\, \left[-\frac{\pi}{6}\, \Big(65\, \delta\,
      \chi_{a} + (65-68\,\nu)\, \chi_{s} \Big) \right] \right\}.
  %%% PLEASE KEEP THIS EQUATION FORMATTED AS IS!!!
  %%% It is designed to blend with the given reference.
\end{equation}
Though formally high in order, this term is large enough in magnitude
(for the spins we present) to dominate the next-to-leading-order spin
contribution to the flux during most of the simulations, dominating
even the leading-order term several hundred $M$ before merger.

The orbital phase is computed using an energy-balance equation
incorporating the rate of change of orbital energy and the loss of
that energy in the form of tidal heating and the gravitational-wave
energy flux $\mathcal{F}$.  Various methods exist---referred to as the
TaylorT1, T2, T3, and T4 approximants~\cite{Damour2001,
  Buonanno-Cook-Pretorius:2007, Boyle2007}---for integrating these
equations, all of which should be equivalent to the \pN order
available, in the sense that they differ only by higher-order \pN
terms.  This suggests that all four approximants should agree with
each other and with the numerical waveform, within the uncertainty of
the \pN approximations.  We have re-derived each of these approximants
using the expressions described above and compared the results of each
to the numerical waveforms.

\begin{figure}
  \includegraphics[width=0.495\linewidth]%
  {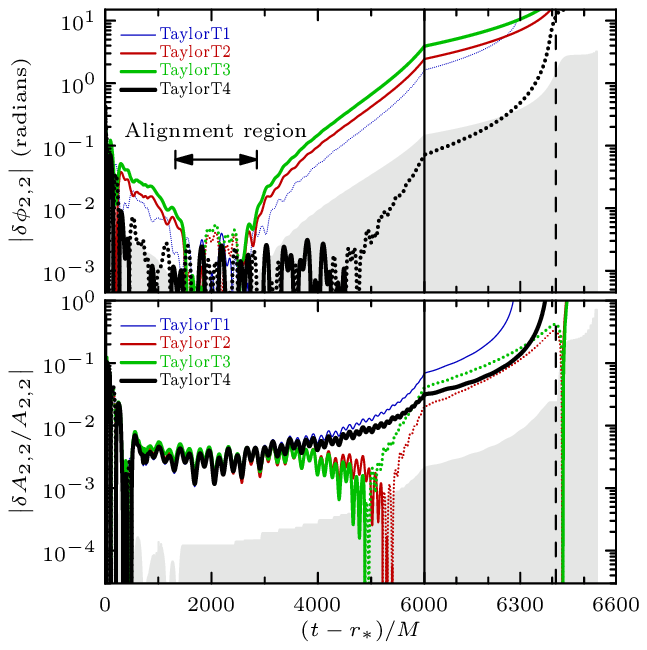}
  \hfil
  \includegraphics[width=0.495\linewidth]%
  {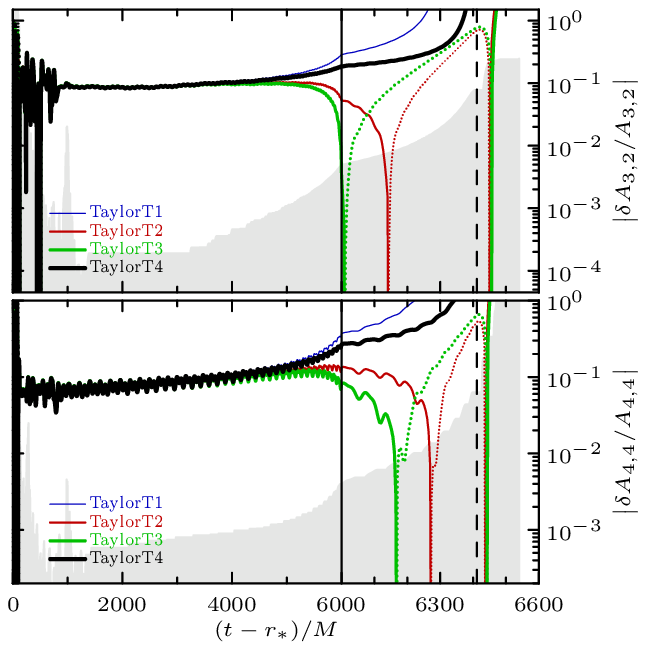}
  \caption{ \label{fig:Align97PN} %
    Comparison between numerical and \pN data for the $\Aligned$
    system for the four \pN approximants.  \textit{Left panel:} Phase
    and relative amplitude errors for the dominant $(\ell,m) = (2,2)$
    mode.  \textit{Right panel:} Relative amplitude errors for the
    next two leading modes, $(\ell,m) = (3,2)$ and $(4,4)$.  In all
    plots, the shaded gray region denotes our estimate for the
    numerical uncertainty, described by
    equation~\ref{eq:WaveformUncertainty}.  Finely dotted portions of
    the graphs indicate negative errors; the \pN quantity is larger
    than the numerical quantity in these regions.  Note that the scale
    on the horizontal axis changes at $(t-\rstar)/M = 6000$ in each
    plot for improved visibility of the merger and ringdown. %
  }
\end{figure}

\begin{figure}
  \includegraphics[width=0.495\linewidth]%
  {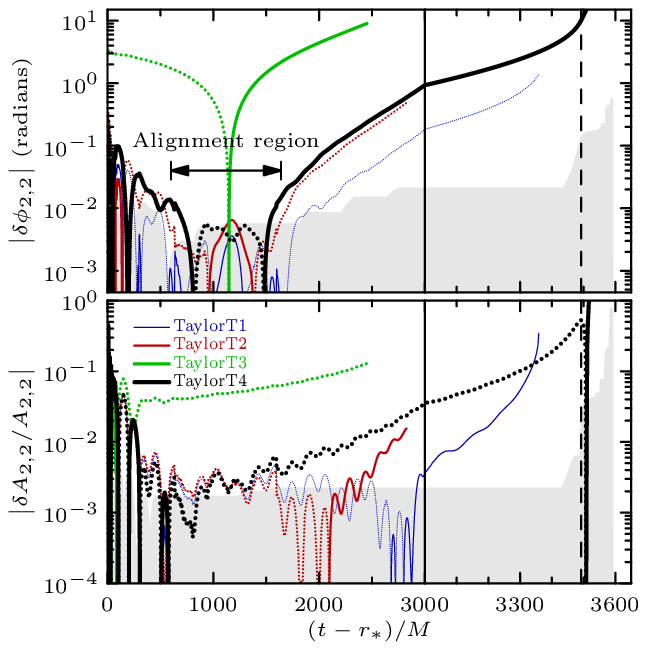}
  \hfil
  \includegraphics[width=0.495\linewidth]%
  {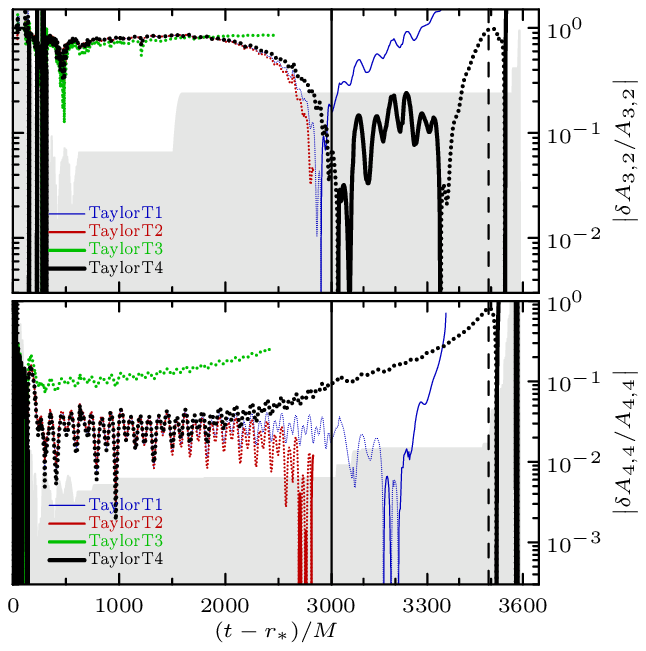}
  \caption{ \label{fig:Anti95PN} %
    Comparison between numerical and \pN data
    as in figure~\ref{fig:Align97PN} but for the $\AntiAligned$
    system.  Note the difference in horizontal scale compared to
    figure~\ref{fig:Align97PN}.  Note that the scale on the horizontal
    axis changes at $(t-\rstar)/M = 3000$ in each plot for improved
    visibility of the merger and ringdown.  %
  }
\end{figure}

Figures~\ref{fig:Align97PN} and~\ref{fig:Anti95PN} show the \pN
comparisons for $\Aligned$ and $\AntiAligned$, respectively.  Only the
phase of the $(2,2)$ mode is shown, because the phases of other modes
are integer multiples of this to a good degree of accuracy.  The gray
region in the background of each plot shows our uncertainty estimate
for the numerical data of the given quantity, given by
equation~\eref{eq:WaveformUncertainty}, and made to be a
non-decreasing function of time after the beginning of the alignment
region.

One remarkable feature is that the TaylorT4 approximant captures the
phase surprisingly well for the $\Aligned$ system (black line in the
top left plot of figure~\ref{fig:Align97PN}); it agrees with the
numerical data within the uncertainty for roughly $3400M$ after the
end of the alignment region (in which it is \emph{forced} to agree
with the numerical data).  This brings it within $200M$ of the merger.
Contrast that agreement with the other approximants, which disagree
with the numerical data immediately after (or even before) the end of
the alignment region.

Of course, this agreement of TaylorT4 is presumably pure coincidence,
as all approximants agree with each other within the uncertainty of
the \pN approximations.  The same coincidence was found in the
equal-mass nonspinning case~\cite{Boyle2007}, but has been shown not
to carry over to systems with other parameters (see,
e.g.,~\cite{Hannam2007c}).  Indeed, looking at the $\AntiAligned$
system (figure~\ref{fig:Anti95PN}), we see that the TaylorT4
approximant is actually the second worst in that case---the sole worse
approximant being TaylorT3, which has \emph{decreasing} orbital
frequency starting just before the simulation, and is thus an
especially poor description of the waveform.\footnote{To be more
  precise, the TaylorT3 approximant expresses the orbital frequency of
  the binary as a function of the \pN time to coalescence by inverting
  a power series (which is used directly for the TaylorT2
  approximant).  That frequency never reaches the initial frequency of
  our simulation; the frequency plot ``turns over'' before that point.
  This is not unusual behavior for TaylorT3.  For example, even in the
  nonspinning case, similar behavior is seen for mass ratio $10:1$.
  Evidently, the series-inversion procedure is not particularly
  accurate.}  We also point out that the $\Aligned$ system inspirals
for roughly twice as long as the $\AntiAligned$ system and reaches
nearly twice the frequency, making the phase coherence of TaylorT4 all
the more surprising.

The figures also show the accuracy of the \pN approximations for the
amplitudes of the three dominant modes.  In each case, the amplitude
of the $(\ell, m) = (2,2)$ mode is the most accurate, becoming worse
for higher modes.  This is to be expected, as the relative \pN order
to which the amplitudes are known decreases with increasing $\ell$.
In particular, the $(2,2)$ mode is known to relative 3-\pN order,
while the $(3,2)$ and $(4,4)$ modes are only known to relative 2-\pN
order.  Nonetheless, because of their far smaller magnitude, these
higher-order modes actually have comparable \emph{absolute} accuracy.

The $(3,2)$ mode is particularly interesting.  In the $\Aligned$ case,
its amplitude is comparable to that of the $(4,4)$ mode.  However, in
the $\AntiAligned$ case, the $(3,2)$ mode is far smaller until the
merger.  For most of the inspiral, the \pN amplitude error is very
large---being off by roughly 80\%.  Again, however, this error is
relative; the \pN approximation correctly predicts that the amplitude
should be quite small in this case, because of a cancellation between
the leading-order nonspinning and spinning components of the
amplitude.

Finally, we note that both waveforms can be hybridized to \pN
waveforms at frequencies of $M\, \omega \approx 0.035$, though these
hybrids are not necessarily accurate enough to be useful in parameter
estimation for detector-data analysis.  As in
reference~\cite{Boyle:2011dy}, we can estimate the error in any hybrid
by measuring the mismatch~\cite{SathyaprakashDhurandhar:1991,
  BalasubramanianEtAl:1996, Owen:1996} between each pair of hybrids
formed with the various approximants TaylorT1--T4; the error estimate
is the maximum such mismatch.  Using the Advanced LIGO high-power
noise curve with no detuning~\cite{Shoemaker2009} to do this
measurement for the $\Aligned$ system, we find mismatches larger than
$0.01$ for total masses below roughly $40\, \MSun$.  This means that,
for any detected ($\text{SNR} \gtrsim 8$) system with a lower mass,
the uncertainty in these hybrids would be larger than the statistical
uncertainty due to noise in the detector~\cite{Flanagan1998a,
  Lindblom2008}.  For stronger signals or lower masses, more accurate
\pN waveforms and/or longer numerical simulations would be needed.
For the $\AntiAligned$ case, a similar comparison would lead us to
conclude that the hybrid is completely uncertain because of the bad
behavior of the TaylorT3 approximant.  If, on the other hand, we
exclude that approximant as anomalously bad, we find that the hybrids
are only accurate enough for parameter estimation above roughly $60\,
\MSun$.  Still, it is possible that such hybrids would be accurate
enough for \emph{detection} purposes~\cite{OhmeEtAl:2011}.

\section{Conclusion}
\label{sec:Conclusion}
The simulations discussed in this paper contain the most nearly
extremal BBHs simulated to date. In our spin $0.97$ simulation, we
have found remarkably good agreement between the horizons' mass and
spin evolutions and Alvi's analytic predictions, but we have found
only moderately good agreement between the remnant hole's final spin
and several analytic formulas, which suggests that these analytic
formulas could be improved significantly using a set aligned and
anti-aligned nearly-extremal BBH simulations.

We have found that the waveform from the 12.5-orbit anti-aligned case
$\AntiAligned$ exceeds the NINJA-2 accuracy requirements, while the
waveform from the 25.5-orbit aligned case $\Aligned$ exceeds the
NINJA-2 amplitude requirement but (because it is so long) fails to
meet the NINJA-2 phase requirement (although it does meet the phase
requirement easily if truncated to 5 orbits, the NINJA-2 length
requirement).

These results demonstrate the feasibility of applying waveforms from
numerical simulations to gravitational-wave data analysis efforts when
the holes have nearly extremal spins---a case previously inaccessible
numerically but relevant astrophysically, given the evidence that
nearly extremal black holes could exist. For example, waveforms such
as those considered in this paper could be used in calibrating
analytic template banks used for gravitational-wave detection
searches. To pursue this goal, we plan to apply our methods for
evolving nearly extremal BBHs to a large variety of BBH
configurations, including unequal masses and spin precession.

We have compared our numerical waveforms to several $\pN$
approximants, finding that the $\pN$ and numerical waveforms disagree
at times well before merger.  We also find that the $\pN$ approximants
disagree with one another, indicating a large uncertainty in the $\pN$
approximations which leads to a large uncertainty in the resulting
hybridized waveforms.  Extracting the BBH parameters from detector
data for systems with nearly extremal spins will require far longer
numerical simulations, far more accurate $\pN$ waveforms, or a
combination of the two.  In the absence of improved $\pN$ waveforms,
however, this implies that parameter estimation when the holes have
nearly extremal spins could prove quite challenging, since the longer
numerical simulations that would be necessary will come at high
computational cost.

% \begin{acknowledgments}
\ack We are pleased to thank Nick Taylor for a gauge modification that
allows us to use the non-smooth maps of
reference~\cite{Szilagyi:2009qz} throughout our evolutions and Larry
Kidder, Robert Owen, Harald Pfeiffer, Saul Teukolsky, and Kip Thorne
for helpful discussions.  This work was supported in part by grants
from the Sherman Fairchild Foundation to Caltech and Cornell and from
the Brinson Foundation to Caltech; by NSF Grants No. PHY-0601459,
No. PHY-1068881, and No. PHY-1005655 at Caltech; by NASA Grant
No. NNX09AF97G at Caltech; by NSF Grants No. PHY-0969111 and
No. PHY-1005426 at Cornell; and by NASA Grant No. NNX09AF96G at
Cornell.  The numerical computations presented in this paper were
performed primarily on the Caltech compute cluster \textsc{zwicky},
which was funded by the Sherman Fairchild Foundation and the NSF
MRI-R\textsuperscript{2} grant No. PHY-0960291 to Caltech.
\MarkComment{Maybe leave out the MRI-R stuff and have the grant number
  instead, or should we have both?}  Some computations were also
performed on the GPC supercomputer at the SciNet HPC Consortium;
SciNet is funded by: the Canada Foundation for Innovation under the
auspices of Compute Canada; the Government of Ontario; Ontario
Research Fund - Research Excellence; and the University of Toronto.
Some computations were performed in part using TeraGrid resources
provided by NCSA's Ranger cluster under Grant No. TG-PHY990007N.
% \end{acknowledgments}

% \appendix
% \section{Some Appendix}

%%%%%%%%%%%%%%%%%%%%%%%%%%%%%%%%%%%%%%%%%%%%%%%%%%%%%%%%%%%%%%%%
\section*{References}
\bibliographystyle{unsrt} \bibliography{References/References}
\end{document}